# Keeping Community in the Loop: Understanding Wikipedia Stakeholder Values for Machine Learning-Based Systems

C. Estelle Smith[1], Bowen Yu[1], Anjali Srivastava[1], Aaron Halfaker[2], Loren Terveen[1], Haiyi Zhu[3]
[1]University of Minnesota, [2]Wikimedia Foundation, [3]Carnegie Mellon University
smit3694@umn.edu, bowen-yu@umn.edu, anjali@umn.edu, ahalfaker@wikimedia.org,
terveen@umn.edu, haiyiz@cs.cmu.edu

## ABSTRACT
On Wikipedia, sophisticated algorithmic tools are used to assess the quality of edits and take corrective actions. However, algorithms can fail to solve the problems they were designed for if they conflict with the values of communities who use them. In this study, we take a Value-Sensitive Algorithm Design approach to understanding a community-created and -maintained machine learning-based algorithm called the Objective Revision Evaluation System (ORES)—a quality prediction system used in numerous Wikipedia applications and contexts. Five major values converged across stakeholder groups that ORES (and its dependent applications) should: (1) reduce the effort of community maintenance, (2) maintain human judgement as the final authority, (3) support differing peoples' differing workflows, (4) encourage positive engagement with diverse editor groups, and (5) establish trustworthiness of people and algorithms within the community. We reveal tensions between these values and discuss implications for future research to improve algorithms like ORES.

## INTRODUCTION
Automated and artificially intelligent algorithmic systems are being used to govern digital worlds. For example, on Facebook, algorithms manipulate the order of the posts users see on their news feed, and help identify and censor trolls, fake news, terrorism, and racist or sexist ads [55]. Ride-hailing apps like Uber, Lyft and DiDi rely on intelligent algorithms to automatically optimize and assign tasks to workers. On Wikipedia, sophisticated algorithmic tools are used to assess the quality of edits and to take appropriate actions such as reverting problematic edits [34]. This trend is also reflected offline, as algorithms are increasingly being used to make important decisions and manage human activities. Examples include: matching students into schools [10], helping judges decide whether defendants awaiting trial should be detained or released [11], and helping employers filter resumes [58].

However, sophisticated algorithms that optimize standard accuracy measures can fail to solve the problems they were designed to tackle when they are inconsistent with—or even harm—important values and needs of the people and communities who use them. As an online example, although Wikipedia's quality assessment systems can efficiently detect and revert low quality edits, research shows that they can also harm the motivation of well-meaning newcomers, still learning how to contribute [30]. When their first few edits were rudely reverted by algorithmic tools, Wikipedia newcomers left in droves [33], violating the community's "don't bite the newcomers" policy [1]. Unfortunately, low newcomer retention has hindered the overall growth of Wikipedia [54].

A second example demonstrates that algorithmic innovation can also be highly controversial in the offline world. Without engaging its citizenry, the county, city, and public schools of St. Paul, Minnesota created an agreement to use students' information to predict children at high risk for juvenile delinquency [38]. Public outcry from the community detailed concerns such as potential bias against children of color, thus the "Cradle to Prison" algorithm was formally put to rest.

These two examples show that even technically sound algorithms may harm stakeholders' values, or be rejected outright. To address problems like these, Zhu et al. proposed the method of "Value-Sensitive Algorithm Design" (VSAD), which emphasizes the importance of uncovering a wide range of stakeholders' values at an early stage of the design process, and incorporating and balancing stakeholders' values in the creation of the algorithm [69]. "Value" here is defined as *"what a person or group of people consider important in life"* [6]. VSAD uses bottom-up design: starting by conceptualizing community stakeholders' values, and then using that understanding to guide the design choice of algorithms.

In this paper, we describe the approach and results of an interview study with stakeholders in a machine learning-based quality prediction system on Wikipedia called the **Objective

Revision Evaluation Service (ORES). Our results identify five **Convergent Community Values (CCVs)**, and two additional values from ORES' creator. Our results also identify critical tensions and trade-offs between these values. Our work raises critical considerations for the future re-design of ORES, while contributing to a broader understanding of human values related to Artificial Intelligence (AI) in online communities.

## RELATED LITERATURE
A growing body of work aims to incorporate human values, including fairness, accountability, and interpretability, into algorithm design. However, this attention has been distributed across disciplinary literatures, including machine learning (ML) and Human-Computer Interaction (HCI).

### Fairness, Accountability, Interpretability in ML
There are now two major conferences, FAT* and AIES, devoted to work on fairness, accountability and interpretability in algorithm design. Most relevant and well-known are works on fairness-aware and interpretable machine learning (see [9] and [16] for surveys of these two areas). In particular, much of the fairness-aware machine learning research aims to build predictive models that satisfy fairness notions that are formalized as algorithmic constraints, including statistical parity [17], equalized opportunity [35], and calibration [49]. For many of these measures, researchers have explored a range of trade-offs, such as fairness and accuracy [3, 41, 45]. Three main techniques are used to interpret a trained machine learning model: sensitivity or gradient-based analysis [51, 43], building mimic models [37], and investigation on hidden layers [5]. However, Veale et al pointed out that these approaches and tools are often built in isolation both from specific users and user contexts [56]. Instead of collecting design evidence from real users, this line of research tends to be built on researchers' intuitions of ways to explain and interpret the algorithms [46].

### HCI Research on Algorithmic systems
On the other hand, HCI researchers have conducted surveys, interviews, or analysis on public tweets to understand how real-world users perceive and adapt to algorithmic systems [14, 15, 18, 20]. For example, Lee et al. conducted some interesting qualitative work to understand how humans make sense of and deal with algorithmic management, and how different stakeholders hold varying notions of "fairness" [44]. Yet it remains unclear how to translate these empirical understandings into algorithm design.

There exists a gap and opportunity to bridge these two research areas, drawing principles and methods from Human-Computer Interaction to meet both human and community design needs for algorithmic systems. Value-Sensitive Algorithm Design (VSAD) is one approach that attempts to fill this gap [69].

### Value-Sensitive Algorithm Design
HCI researchers have long argued that human values should be considered in a principled and comprehensive manner throughout the process of technology design [22]. For example, Value-Sensitive Design (VSD) is a tripartite methodology, consisting of iteratively applied conceptual, empirical, and technical investigations. The goal is to prevent biases in design choices or compromises of important user values [6]. Moreover, Zhu et al. proposed the VSAD design framework, which focuses on incorporating human values in the design of algorithms and algorithmic tools [69]. Their approach emphasizes: (1) working closely with relevant stakeholders to uncover their values in the early creation of algorithms; (2) using their values to guide specific design choices at various stages of algorithm design; and (3) designing and evaluating algorithms based not only on accuracy, but also on acceptability and broader impacts.

Although Friedman and Kahn proposed a list of "universal" values[1] [22], Borning and Muller suggest that empirical investigation is important to: 1) contextualize this list with information about how the values play out in a given context of interest, including who held the values, in what milieu, and for what purpose; and 2) uncover values that were not previously seen in publications [6]. Adopting the views of Borning and Muller and Zhu et al., this work provides an empirical study of stakeholder values in the sophisticated Wikipedia context.

## BACKGROUND INFORMATION
### Algorithmic Governance of Wikipedia Communities
Wikipedia is a large scale peer production project with an open governance system [19] and an equally open system for developing/deploying algorithmic support for Wikipedian activities [24, 34]. Despite the basic strangeness of Wikipedia's organizational structure when compared to traditional media, the online encyclopedia has been extremely successful and widespread [40], prompting researchers to question issues of fairness and representation (e.g., [36]).

Due to its success and the open nature of social and algorithmic governance, Wikipedia is a fascinating context to explore questions of agency, distributed governance, and the role of algorithms in social processes. For example, processes like requesting an edit to an article or deleting a problematic article exist specifically because bots enable new workflows that are embedded in complex social contexts [24, 25]. Thus, a rich and growing literature explores the social [23] and governance roles [48] that bots have taken on in Wikipedia.

Often the "hard laws" of a bot's code can come into conflict with the "softer," contextual, social rules of humans. This has lead Wikipedians to ask new types of questions about how bots and other algorithmic support tools should behave [23]. The "laws" of bots can even manifest as conflict between human bot developers [28]. Wikipedia's own algorithmic support for quality control has been implicated as the cause of many of Wikipedia's problems with newcomer socialization [30]. Due to the nature of AIs as a complex and opaque technology, Wikimedia staff have sought to build novel types of transparency in their machine prediction service for Wikipedians, ORES [31]. In this study, we explore values and expectations related to Wikipedians' use of ORES.

### Objective Revision Evaluation Service (ORES)
ORES is a web service and application programming interface that provides real-time predictions on edit quality and article

---
[1] Including: Human Welfare, Privacy, Freedom From Bias, Universal Usability, Trust, Autonomy, Informed Consent, Accountability, Identity, Calmness, Environmental Sustainability, Ownership & Property

quality [60]. The system was originally developed and deployed in order to make machine prediction technology more available to volunteer tool developers who support quality control processes on Wikipedia. By providing open access to transparently developed machine prediction services, the developers of ORES intend to enable a broader set of stakeholders to become more engaged in developing processes/technology to support quality control work on Wikipedia [30, 31].

Supported by an engineering team at the Wikimedia Foundation [65], ORES has been online since July 2015 and currently supports 42 different languages of Wikipedia (e.g., English, Bengali, Arabic, German, Swahili, etc.). ORES has become the underpinning of an entire suite of tools to support quality control, newcomer socialization, and task routing [61, 29].

### ORES Stakeholder Groups

Open peer-produced and peer-governed communities like Wikipedia continually evolve and re-invent themselves. Community members—both within and external to the Wikimedia Foundation (WMF)—often have multiple roles that are dynamic and tend to shift over time. Here, we describe key ORES stakeholder roles, which occasionally overlap.

**Tool Developers.** Among Wikipedia's volunteer community are a class of contributors referred to as tool developers. These software developers build bespoke software support systems for Wikipedian's processes such as tools for detecting and reverting vandalism, organizing task lists, and managing a queue-based work process [24]. These tools come in the form of robots that automatically edit wiki pages (e.g. SuggestBot [64, 12]), javascript "gadgets" (e.g. Real-time Recentchanges [68]), and stand-alone UIs outside of Wikipedia's regular browser (e.g. Huggle [67]). Tool developers use ORES to augment their tools and provide useful quality predictions to end-users.

**Wikimedia Product Teams.** WMF maintains product teams tasked with building important software tools to support Wikipedia's editors and readers. They are responsible for adding features to the wikipedia.org property as well as to the official Android and iOS applications. These teams have used ORES in features for making counter-vandalism work easier [59] and for making new page patrolling easier [62].

**Editors.** ORES predictions are not directly delivered to end-users (i.e. reviewers and editors on Wikipedia). Instead, they are mediated indirectly by tools and interfaces maintained by volunteer developers and Wikimedia product teams. Although Wikipedia editors are not directly interacting with ORES, the source and nature of those predictions remain of interest to them. In quality control work (e.g. counter-vandalism), both the reviewers and the reviewed users have a stake in how ORES works. For instance, a false positive error by ORES would be a high quality edit that was errantly predicted to be damaging. If ORES has a high false positive rate, more users completing good edits will suffer. Similarly, reviewers who use ORES predictions will value a high level of precision so that they don't need to spend most of their time reviewing false positives [27]. It is often these end-users who raise issues of bias or inaccuracy with ORES' maintenance team.

**Researchers.** Scholarly researchers use ORES in many ways. Dang et al. used ORES models and training data as a baseline for experimenting with novel modeling strategies [13]. Halfaker used ORES quality models to measure the rise and fall of specific coverage gaps in Wikipedia [29]. Vincent et al. used ORES to filter low quality content from English Wikipedia when exploring usage on other sites such as Stack Overflow and Reddit [57]. As in this study, researchers are interested in understanding how ORES affects various Wikipedia communities, and effective strategies for governance and control of the algorithms through Wikipedian governance mechanisms.

## METHODS

In this study, we employ a grounded theory approach to understand the values of ORES stakeholders. We first posted our study on a research meta-wiki [63] and elicited community feedback to ensure that our study will not disrupt the mission of Wikipedia [2]. We then conducted semi-structured interviews [53] with 16 participants over phone or video chat. We asked participants about their stakeholder role and experiences related to ORES or ORES-related applications, and their opinions and ideas about future ORES development efforts. Interviews were transcribed, coded, and affinity mapped in immersive group meetings. We discussed and iterated on themes as they emerged, and recruited until we achieved data saturation for each stakeholder category. Because the Wikipedia community values attribution, we provided all participants the option to self-disclose their identity in this paper. Table 1 contains voluntarily disclosed participant information.

### Participant Recruitment and Relationship to Authors

As a Principle Research Scientist at WMF, P1 is primarily responsible for the initial development and ongoing maintenance of ORES. Also listed as an author of this study, P1 participated in framing and writing the paper, and recruiting by connecting our research team to WMF employees, external researchers, and developers. Our research team recruited additional editors by posting on Village Pump, the Teahouse Talk page, and random editors' talk pages. As ORES' originator, we felt it necessary to include P1's perspective in our data; therefore he was also interviewed by the first author as a participant.

In autoethnography, participant observation, and action research methods, authors are occasionally participants in their work. For example, Geiger's recent works feature vignettes gleaned from over a decade of personal involvement in Wikipedia [24, 25]. While P1's data strengthens our study by providing context that no one else could have, we did not observe conflicts between his and other participants' data. Even if we excluded his interview, w are confident that the contributions of this work would be similar. However, we acknowledge that this methodology shares the strengths and limitations of other authoethnographic work; i.e. we gain in-depth perspective from a critical stakeholder with arguably the most extensive knowledge of ORES, yet we may simultaneously risk amplifying his biases and influence.

To offset this limitation, P1's data was acquired/analyzed in the same manner as all other participants. We excluded P1 from our analytical protocol, and P1 did not contribute to writing the results. More closely resembling Glaser's recommended methodology of Grounded Theory than Corbin and Strauss, the first author had only limited knowledge in this domain at the start of data gathering. Themes emerged organically during analysis, and were later supplemented by existing work. These measures, though imperfect, reduce bias in presentation of the results, while retaining invaluable data. As a validity check, we offered the Wikipedia community to weigh in on our interpretation. We posted a preliminary draft to the Village Pump forum [66]. We informed participants of this post and incorporated minor revisions based on all feedback received.

While we invested significant effort to recruit across the spectrum of ORES stakeholders, we acknowledge the additional limitations that: (1) while interviews provide valuable in-depth data, they only capture perspectives from a small number of participants, and (2) our participants may not perfectly represent all members of the community. It is especially challenging to recruit newcomers and people who aren't familiar with how Wikipedia operates. Future work should engage larger numbers of editors about their values, potentially through large scale, lightweight, contextual surveys.

**RESULTS**

In our results, we first discuss **creator values**. Apart from some backend engineering, most ORES models and its framework were designed and built entirely by P1, who embedded core values into ORES' first instantiation that remain influential to community usages of ORES. Second, although we initially expected that different stakeholder groups might emphasize different values, we were surprised to find no obvious conflicts between *stakeholder groups*. However various tensions exist between different *values* themselves. Therefore, we refer to stakeholder values broadly as **Convergent Community Values (CCVs)**. Throughout the paper, we will refer to the five CCVs that emerged through our analysis by these abbreviated phrases:

1. **Effort Reduction:** Reduce the effort of community maintenance.

2. **Human Authority:** Maintain human judgement as the final authority.

3. **Workflow Support:** Support differing peoples' differing workflows.

4. **Positive Engagement:** Encourage positive engagement with diverse editor groups.

5. **Community Trust:** Establish the trustworthiness of people and algorithms within the community.

Furthermore, throughout each section, we provide **bolded recommendations** for specific ways in which that value can be operationalized by system designers and researchers to guide the design of algorithmic systems (including but not limited to ORES) and social structures surrounding them. We conclude results by reporting tensions in enacting these creator values and CCVs in ORES-related development efforts.

**Creator Values**

*Enable Consistency and Replicability*

Prior to ORES, each of the anti-vandalism and quality control bots and tools on Wikipedia were supported by their own internal models. In particular, ORES' direct predecessor, Sticky, was a machine-learning model that had been built to support a tool called Huggle[2] by providing predictions of edit quality. P1 said, *"Sticky's API was great. It allowed me to implement Snuggle[3], but it was deeply broken and useless for many types of re-appropriation"* because its scores were only available in a temporary cache; those scores could not later be accessed or re-created. As a researcher, P1 wanted to be able to generate scores for both current and historical edits. In order to replicate experiments or analyses later or differently, those scores needed to be permanently available.
**[Recommendation 1:] Developers should build affordances that make the outputs of algorithmic systems permanently available and replicable.**

*Facilitate Experimentation*

P1 stated that ORES is intended to function as an infrastructure that facilitates *"an explosion of experimentation and change."* P1 wanted to allow others to also re-appropriate and study open source algorithms. *"What I really hoped to see wasn't that we would do quality control better, exactly, but that more people would start experimenting with quality control tools."* **[Recommendation 2:] Existing algorithmic systems should be made available to members of open source communities for re-appropriation and experimentation.**

P1's motivation in creating ORES resonates another WMF employee who described the need for an even higher level infrastructure for experimentation—one that enables anyone to influence algorithms. *"We need to build a contribution platform that allows people to plug their own algorithms in." (P4)*
**[Recommendation 3:] Open source communities should develop platforms that enable members to implement and deploy new algorithmic systems within them.**

**Convergent Community Values**

Set against the preceding creator values, we next describe five CCVs that may guide development efforts for machine-learning based algorithmic systems on Wikipedia.

*Reduce the Effort of Community Maintenance*

Given ORES' function as a generator of quality predictions, participants agreed that ORES-dependent tools or systems need to help reduce the sheer volume of human labor necessary to maintain the online encyclopedia. Wikipedia's content and readership continue to grow, even as its communities of active editors do not. *"We have to augment our capacity, and well-designed algorithms put to the right purposes are a key component in that." (P3)*

According to community, the "right purposes" are filtering, recommendation, and prioritization of content or tasks that help to focus human efforts in quality control workflows. For example, P4 said that editors need to be able to *"sort the queue and decide to spend their time on the things that deserve it."* Similarly, D1 said that ORES' role is to *"help focus the efforts of the reviewers. When there's a lot of edits in a particular minute, they cannot review them all."* **[Recommendation 4:] When it is impossible to review all issues, algorithmic tools need to surface and emphasize those issues that are *most critical* to be examined by human eyes.**

---
[2] A quality control tool that helps experienced Wikipedians look for bad edits in order to revert them and maintain article quality.
[3] A newcomer socialization tool that helps experienced Wikipedians look for good edits by good editors, and invites them to send newcomers welcome messages or invitations to help spaces.

| ID | Role Description | Registered since | # Edits | Gender | Handle | Full Name |
|---|---|---|---|---|---|---|
| P1 | WMF Principal Research Scientist, ORES Creator | 2008 | 4k | M | EpochFail | Aaron Halfaker |
| P2 | Intern, Helped Implement ORES | - | - | - | - | - |
| P3 | WMF Senior Researcher, Teahouse Founder, HostBot operator | 2008 | 5.6k | M | Jtmorgan | Jonathan T. Morgan |
| P4 | WMF Product Manager | 2017 | 1k | M | - | - |
| P5 | WMF Director of Engineering | 2004 | 63k | X | - | - |
| R1 | External Researcher | 2004 | - | F | - | - |
| R2 | External Researcher | - | - | M | - | Ofer Arazy |
| D1 | Volunteer App Developer | 2009 | 3.5k | - | Krinkle | Timo Tijhof |
| D2 | Wikimedian, Volunteer PAWS* Maintainer | 2006 | 21.8k | M | Chicocvenancio | Chico Venancio |
| E1 | Editor, Admin | 2007 | 156k | F | Rosiestep | Rosie Stephenson-Goodknight |
| E2 | New Page Patroller (NPP), Admin | 2005 | 13.5k | M | Barkeep49 | - |
| E3 | Editor, Teahouse Host | 2010 | 40k | M | Nick Moyes | Nick Moyes |
| E4 | Editor, Teahouse Host, AfD**, Guild of Copy Editors | 2010 | 10k | M | Timtempleton | Tim Templeton |
| E5 | Editor | 2018 | 1k | M | - | - |
| E6 | Editor, Vandalism Reverter, NPP | 2018 | 17.5k | M | SkyGazer 512 | - |
| E7 | Editor | 2018 | ∼300 | M | Ohanwe Emmanuel .I. | - |

Table 1. Participant Summary. Participants were offered the choice to voluntarily disclose their personal information. A single dash indicates that they did not provide that information. * *PAWS is a web interface for jupyter notebooks (https://www.mediawiki.org/wiki/PAWS)*. ** *Articles for Deletion*.

Developers described an additional need to reduce editor fatigue, esp. in the case of highly backlogged tasks (e.g. fighting vandalism, engaging with new editors, fixing sources, reviewing new articles), and in communities with small and limited subsets of active users. *"If we can leverage the manpower that we do have with more automation...these people will have less backlog and can focus on other contributions."* (D2) **[Recommendation 5:] Algorithmic tools should be designed to assist with those tasks that are especially draining for humans due to continual or overwhelming volume.**

While this first CCV is perhaps the most obvious, it nonetheless demonstrates a successful case in which an algorithm was designed, deployed, and embraced by the community. Encouragingly, editors described using ORES-dependent tools consistently with the goals for which they had been created. For Page Curation, E2 said ORES' predictions *"give me a heads-up of things that I might want to pay particular attention to."* For anti-vandalism in Recent Changes, E5 said that color coding edits based on ORES' predictions *"lets me see the ones I currently need to see rather than the ones that I'm fairly certain aren't needed."*

*Maintain Human Judgement as the Final Authority*
Less obvious than the first CCV—and more debatable—is the concept that humans should retain the final authority in decision-making. Some AIs on Wikipedia are fully automated, e.g. ClueBot, which rapidly and automatically reverts nearly certain cases of vandalism (and is not built on ORES). On the other hand, ORES is typically integrated into tools as a high-level filtering mechanism for a large feed of information, rather than a final decision-maker. Participants expressed a sense of caution towards fully automated processes and emphasized the importance of retaining individual human and even community judgement as the final arbiter. For example, *"I feel like the community is pretty good at and pretty protective of the idea that humans should make the final decisions that affect content."* (P5) **[Recommendation 6:] Human judgement rather than algorithmic output should be the final arbiter on individual decision-making instances.**

Furthermore, AIs can be good and useful, yet must always be monitorable by the community. *"If we could have a very accurate bot based on ORES, good...but it would always be subject to community concerns, someone having a 'stop' button on the bot."* (D2) **[Recommendation 7:] Algorithmic systems should be continually monitorable by humans after being deployed.**

ORES provides users with the ability to select prediction confidence thresholds. Interestingly, editors agreed that AIs built on ORES should emphasize recall (i.e. flagging more potentially problematic edits) over precision (i.e. ensuring that flagged edits are definitely problematic). E.g.,

> *ORES' purpose is more to create lists of possible problematic pages or edits for human editors to look at, rather than take action fully automatically. If this is the case, there's not a huge obligation to be very restrictive, unlike fully automated processes like ClueBot. (E6)*

*I would err on the side of giving the bot editing leeway, particularly if there are people monitoring them as well to make sure that they're working correctly. (E4)*

**[Recommendation 8:] Human stakeholders should have authority to select confidence thresholds, and determine acceptable trade-offs between key system criteria of an algorithmic system.**

An important but more subtle distinction is that humans must not only have the ability to stop or monitor bots, but also to genuinely rely on their own judgement rather than algorithmic output. For example, E1 has a workflow that involves identifying poor quality articles (stub or start-class), and then editing them until they are upgraded to a C-class article. E1 said, *"I wouldn't rely on ORES 100% of the time. As soon as it would change to a C, then I would stop, but I would still have to use my brain to make a decision."* In other words, this editor's judgement of what a C-class article should look like does not always match ORES, and is also more important than what ORES predicts. **[Recommendation 9:] Humans should be reminded to rely on their own judgment rather than defaulting to accept an algorithm's output.**

*Support Differing Peoples' Differing Workflows*
According to the prior CCV, humans should always have the final say; yet different humans choose different things based on their own preferences. *"You could decide more into this direction, but I could go more into this direction...how do you make these differences also available in ORES?" (R1)*

Our data suggest that tools built on ORES can either support or hinder editors in their workflows. For example, P1 described a problematic workflow supported by Huggle—an ORES-dependent tool that functions like a third-party application. Huggle's interface shifts editors out of the Wiki context to help them identify and revert bad edits, but makes it inconvenient to open a new browser tab and resolve the issue through any action other than reversion. *"If you look at an edit [in Huggle] and you're like, oh, this is almost good. I can just make this change, and it would be fine. [But] there's no 'Edit the Page' button in Huggle." (P1)* Because of this, editors often end up reverting an "almost good" edit, which conflicts with their primary goal of improving article quality.

Editors described additional ways that Huggle conflicts with their goals. For example, E6 no longer uses Huggle because the interface for viewing diffs is challenging to navigate, doesn't show consecutive diffs, and in cases of vandalism, *"it's time consuming to look for previous warnings, and to use a different warning level than what is put automatically."* These data show that if tools obstruct users' workflows and prevent them from achieving specific goals, they risk being abandoned. **[Recommendation 10:] Algorithmic tools should facilitate workflows that help to achieve users' *actual* end goals.**

Furthermore, different editors have different priorities. For example, E3 currently uses the cumbersome procedure of filtering Recent Changes by searching for a set of keywords that he has personally derived to identify topics that he wants to work on. E3 wants ORES filters to improve his workflow by letting him review articles by category. *"I would like to be able to select for me the Recent Changes in a sphere that is important to me and leave somebody else to make the same value judgments as to what's a good and bad edit in a different sphere."* In this vein, P2 shared that he had been experimenting with a topic model for ORES, so that editors *"who are interested in specific topics can look at pages only on those topics."* However, such a model has not yet been deployed.
**[Recommendation 11:] Developers should identify sets of users' priorities throughout their workflows, and build tools that are configurable to those different priorities.**

People also develop different habits for interacting with ORES' predictions, given the affordances of the tools built upon them. For example, in quality prediction applications built on ORES, editors can adjust thresholds using a sliding bar to indicate quality thresholds for either article drafts or individual edits. Some people prefer to help improve possibly good content, e.g., *"I use ORES to filter for drafts that look like they might be good quality so that I can look at those in a timely manner." (P5)* Others prefer to eliminate bad quality content from the queue, e.g. *"[In Recent Changes,] I don't want the 'likely to be okay' edits, so I will go for the worst ones. I think I've got a good setting for me, but I don't know if anybody else would [agree]." (E3)* These data suggest that people often know what they want to look for. However, the sliding thresholds in current ORES-based applications do not communicate these types of goals in an intuitive way. E3 suggests, *"I'd say you can almost do with a nice big button that just says, 'Do you want to see only the good edits? Only the bad edits? Only the middle ones?' And make it slightly easier to understand."*
**[Recommendation 12:] Developers should create intuitive UI/UX elements that make it easy to select workflows based on users' different priorities.**

D1 also shared the interesting observation that editors may switch between different sets of habits or "modes". For instance, one ORES filter distinguishes between good vs. bad faith edits. One use case for this filter is quickly finding and reverting bad edits completed in bad faith, while a distinct second use case is finding low quality but good faith edits and investing more time to help such editors improve the quality of their work. *"You can be in a different mode at different days...Switching back and forth between the two [types of tasks] in the same mode is very difficult,"* and can also result in worse outcomes—e.g. inadvertently being impatient with good faith editors. D1 also pointed out that the distinction between these two use cases is important but not obvious, especially to newcomers, since the UI provides no guidance about the different ways one might use ORES' filters.
**[Recommendation 13:] UI/UX elements in algorithmic tools should be designed to give users the flexibility to select and stay focused on the type of use case they want to work on, until they decide to switch to a different one.**

*Encourage Positive Engagement with Diverse Editor Groups*
The Wikipedia literature has described how quality control processes can have adverse impacts on newcomers [33]. Interestingly, our participants suggest that algorithms like ORES can and should play a role in *helping* inexperienced editors, as well as other underrepresented editor groups, such as minori-ties and females.[4] *"It's very important to allow diversity and to allow very low obstacles and allow as much participation as you can. I think that the article quality is driven to a large extent by the diversity of hundreds of users." (R2)*

Beyond issues with their edits being reverted (causing them to leave the project), P4 explained how Wikipedia has become like an ecosystem, in which certain kinds of people are quite well-adapted. However, *"that limits the diversity of the contributors. So the ecosystem needs to change in order to be more welcoming to certain kinds of people." (P4)* Participants offered guiding insights regarding how to best facilitate community "evolution" by designing effective ways for newcomers to interact with algorithms.

First, newcomers ought to be able to understand when it was an algorithm that has executed an action (e.g. edit reversion) that is likely to be experienced negatively. *"When you get reverted by some kind of algorithm, make sure, I, the new editor, know that it was an algorithm, I know that the algorithm has failings, and that I have a way to appeal that and contact a human to say that I think the thing is wrong." (P4)* Recent work supports this concept, suggesting that *contestability by design* must be required throughout the lifecycle of AI systems in order to protect the rights of data subjects [4]. **[Recommendation 14:] Users should be able to understand which actions were taken by algorithms, which actions were taken by humans, and how to contest decisions.**

Complementing the ability to appeal decisions, R1 said, *"If there is a human involved, you also need a kind of social involvement or social connection that people feel more inclined to stay in the community."* Algorithms should therefore surface the right information to the right people at the right time, as opposed to doing the communication and taking away that human moment. *"Good would be if an algorithm noticed that people were new and then searches for actual humans to write them personalized welcome messages." (P4)*
**[Recommendation 15:] Social connections within the community should be facilitated rather than replaced or weakened by algorithmic systems.**

Mirroring this idea, E7 (a newcomer) was enthusiastic about an algorithm being designed *"so it's kind of empathetic. Like, it's not hostile to new editors, such that maybe, if there is a wrong edit, the algorithm could tell the editor what's the story, and that the edits are not so proper, you could have done it this way."* E7 wants the algorithm to explain *how* he did something wrong, and provide resources for understanding the correct way to achieve goals. **[Recommendation 16:] Algorithmic systems should provide transparent explanations of their behavior, and accessible training resources for effective interactions with them.**

In addition to providing educational resources, D1 explains that editors should be guided to understand how to succeed in the long term as Wikipedia contributors. *"There's not really a path forward to promotion, or to be steered towards what needs to be done."* Editors mirrored D1's interest in such tools. For example, E4 said: *"I signed up for the Guild of Copy Editors. What about an AI tool that actually reaches out to me and says would you be interested in signing up?"* P5 also suggested that an ORES model could *"help some of the admin people really figure out more quickly what's going on with, maybe new editors, or mid-experienced editors, who needed just a boost in something."* I.e. if a particular editor were struggling with a specific rule or conduct issue, an algorithm could surface the problem(s), saving admins the time of tracking them down. **[Recommendation 17:] Algorithmic systems should provide and recommend helpful ways for users to learn and grow within the community.**

Finally, D1 suggests that rewarding editors for good behavior could better serve the long-term goals of Wikipedia. For example, an algorithm like ORES could identify edits that have survived longer or are more stable. *"[Identifying stable edits] could feed back into a model that provides points in some way that does encourage good behavior." (D1)* Prior work validates the perspective that edit stability is a good quality measure via metrics such as Persistent Word Views ("the number of times any given word introduced by an edit is viewed" [50]) and Persistent Word Revisions ("the number of revisions that a word survives" [32]). **[Recommendation 18:] Users should be incentivized by algorithmic systems to behave in ways that create enduring value for the community.**

*Establish the Trustworthiness of People and Algorithms*
Participants agreed that there are opportunities to use algorithms like ORES not only to identify trustworthy people *within* the community, but also to make the algorithm itself more trustworthy *to* the community. P3 described the primary ORES stakeholders as the *"communities that are going to adopt the tool or that are going to consent to having a new bot introduced in their community that does patrol and is built on ORES. I think engaging the community is another way that increases trust, and maintaining trust is important."*
**[Recommendation 19:] Developers should continuously engage with the communities affected by algorithmic systems to build and maintain trust.**

However, not all members of the community are equally trustworthy. Both developers pointed out that Wikipedia currently relies entirely on manual, human processes to establish trust in its community members, since *"Wikipedia [software] doesn't currently have a concept of reputation." (D1)* In the case of positive reputation, some wikis implement higher trust user groups, such as "Bureaucrats" who can grant user group permissions, and "Administrators" who have additional controls over pages and other users' accounts. For example, D1 has "Auto Patrol" status, which allows him to bypass New Page Patrol. *"I've been around long enough, and they trust me that I'm not some vandal who's trying to promote spam."* In the case of negative reputation, D2 explained that reporting users is also an informal, time-consuming process. *"Instead [of reporting users via software], there's a Wiki page with arbitrary content where you create a heading and you say, this is a report. Then you follow some kind of syntax and then you notify someone else."*

---
[4]Due to space constraints, and because newcomers were most frequently discussed, we focus on quotes about newcomers.

On the other hand, E3 described the case of "Request for Adminship", in which theoretically good editors self-nominate to become admins, but must undergo public scrutiny before receiving advanced status. *"I think it would be very good to have a tool that allowed you to analyze all of an editor's edits and pull out and offer you a selection of what it thinks the bad ones are for you to [review]." (E3)* In other words, an algorithm like ORES could help identify editors' historical trustworthy or untrustworthy actions at critical moments of communal scrutiny. P5 expressed concerns that *"the community would not appreciate having every single editor be scored on how trustworthy they are,"* but they also said this concern could be mitigated if people were only scored temporarily (e.g., a 6-month introductory period). To avoid damaging peoples' motivation, such scores should only be accessible to moderators—i.e. not publicly displayed.

**[Recommendation 20:] To aid in community governance efforts, algorithmic systems should provide mechanisms to assess the trustworthiness of community members based on their community contributions and behaviors.**

To build trust in algorithms, Wikipedia editors should also have the ability to affect them. Yet prior work has grappled with the challenge of communicating the esoteric inner workings of algorithms to users (e.g. [8]). Although E3 has little understanding of ORES, he provided a helpful metaphor. *"I don't need to know how a BMW works in order to be able to tell how to drive it."* E3 went on to describe his desire to provide feedback about ORES' performance:

> *I would be in support of human interaction at some state to help strengthen the algorithms. We can give feedback because human editors—that was a wrong edit by ORES or a wrong flag or a right one. But I don't think everybody can do that—like the new page reviewers, you've got to have a number of trusted edits to say we trust this feedback of ORES assessment.*

**[Recommendation 21:]** *Trusted* **users should be able to impact algorithms by providing feedback on their performance, even if they don't understand all details of how the algorithms work.**

**Tensions in Enacting Convergent Community Values**
We found that certain tensions are implicit in enacting the creator values and the CCVs described above.

*Valuing Experimentation vs. Serving CCVs*
P1's goal was to build an infrastructure that enabled experimentation. However, P1 also shared an example of when ORES had been re-appropriated in a manner that damaged the community. A developer built a bot called "Patrobot" for Spanish Wikipedia. Patrobot used ORES raw predictions without considering confidence levels to instantly revert edits more likely to be vandalism than not. However, at its default settings, this model had low precision ( 25%) and was therefore wrong 75% of the time, which is extremely problematic for a bot engaged in auto-reversion. Patrobot was live on Wikipedia for weeks, unable to be turned off by the community. P1 said:

> *You want to set precision, then optimize for recall as opposed to setting a recall and optimizing for precision. ... If somebody doesn't know that documentation is there and they don't have any background in machine classification, then they could certainly just do the same thing again. ... We definitely don't want to put barriers between people using ORES, [but] if you can use ORES, you can also use ORES inappropriately.*

Patrobot effectively violated all five CCVs, yet it was also a reasonable, if naive, attempt at re-appropriation. P4 speaks metaphorically to a broader problem with communicating best practices for experimenting with algorithms:

> *It's like if people became really reliant on Google Translate to communicate with each other, but Google Translate was not perfect in translating and then language degraded over time to become less expressive or something – that would be the concern. So yeah, we would want to encode the norms for using these models into the behaviors that they would take with them.*

This quote highlights the importance not only of clear communication via documentation, but also of embedding norms of use into the design of systems—a concept that has also been raised, for example, in the Reddit context [42].

**[Recommendation 22:] In open source communities, algorithmic system designers need to balance efforts to facilitate experimentation with a strong commitment to community norms and to serve important community values.**

*Positive Engagement vs. Centralization of Power*
Wikipedia represents an attempt to transparently democratize a peer production system, from its code to its content. However, the affordances of algorithmic systems (including but not limited to ORES-dependent applications) creates a situation in which some stakeholders can gain a higher degree of control than others. P1 shared a sense of concern over the balance of power in the system:

> *I think that in a lot of ways, I'm the worst bus factor[5] for ORES...I'm not worried that ORES won't stay online or that people won't keep developing models for it. I do worry that it could be taken out of the control of Wikipedians...I want the people who are on the other side of ORES to have that power and to have less of that power in the hands of people who look like me.*

This quote demonstrates that P1 recognizes his own influence and power, and wants to find ways to ensure that this power is shared more broadly across the community with those who use or are impacted by the algorithms. P1 is currently working on an auditing system for ORES, as a means of enabling diverse editor groups to detect and correct systematic issues with the algorithm. However, another challenge is that many community members may have low algorithmic literacy, and thus require affordances that are less technical. P4 said:

> *People need to have a lot of really specialized knowledge to assemble [algorithms] in the first place, so very*

---
[5]See https://en.wikipedia.org/wiki/Bus_factor

*few people can do it...Those pipes [to contribute to algorithms] need to be even wider and we need to think even harder about how non-technical people can contribute by labeling data, by giving them knobs to tune algorithms.*

**[Recommendation 23:] System designers should vigilantly maintain awareness of their own biases and influence within the community, and strive to make efforts at system development and refinement accessible to *all* community members, including those without specialized knowledge. This can be achieved by providing algorithm end-users ways to understand and tune algorithms and their parameters to better serve community needs.**

*Effort Reduction vs. Positive Engagement*
Many participants shared concerns about algorithmic quality control systems that affect participant motivation. Wikipedia's sophisticated quality control mechanisms can remove bad edits in anywhere from a few seconds to a few minutes [26]. While this high speed is both impressive and desirable with regard to Effort Reduction, D1 worries that the rapid nature of edit reversion can damage the motivation of good faith editors:

*Especially for good faith editors, if they do something, and a few seconds later, they immediately get a notification, your edit is no longer here – maybe those cases can be done later. ... What's the worst thing that will happen? Maybe somebody will see a syntax error for half an hour, an hour, or maybe even a day. That's maybe not so bad. Because we do everything in real time right now, it's very reactive and very combative.*

Whereas individual edits are handled rapidly, many editors described damaging effects to motivation when new article drafts are handled too slowly—e.g., E7 expressed frustration that the work of volunteer editors to create new articles is so often neglected for months on end. *Most importantly, speed up the process of reviewing these articles, so it doesn't take more than a week, because there are a lot of gaps to be filled on Wikipedia. (E7)* This phenomenon is already well understood as a product of the backlogged review tasks that some ORES-dependent applications are intended to solve. Current applications allow editors to organically address edits or articles on unregulated timescales.

**[Recommendation 24:] When designing applications for Effort Reduction, designers should also consider temporality in their designs to be more sensitive to social consequences and not violate the Positive Engagement CCV.**

*Model Fitness vs. Ethical Considerations*
Prior literature in algorithmic fairness has proposed fairness-enhanced predictors and techniques to guarantee fairness of population subgroups under mathematical definitions [21], and further examined tradeoffs between optimizing for model fitness vs. minimizing disparity of different subgroups [3]. In the case of ORES, different groups of people are denoted by their user status as an unregistered or registered user, and if registered, how experienced they are. Most low quality edits come from unregistered and newer users; since ORES uses this as a feature, ORES potentially learns to disadvantage this entire group.

Our data show that there also exist tensions between optimizing model fitness for efficient workflows (which serves the CCV of Effort Reduction) versus what types of features are ethical to include in models. For example, P1 described a design decision in ORES to use only features derived from actual changes to an article, rather than people's reactions to those changes (e.g., warnings), which reduces ORES' precision and recall:

*Even though I think we made the right decision, it still bothers me because it means that when somebody's looking at adopting ORES for quality control work, they could be reviewing half as many edits as they have to review now. I'm essentially almost doubling the amount of work that they do because I don't want to look at rap sheets.*

**[Recommendation 25:] Rather than always selecting optimal models, designers of algorithmic systems should carefully consider the ethics of features they choose to include. If certain features perpetuate systematic biases or disadvantage entire classes of users, it may be more ethical to de-prioritize model fitness, even if that creates more work for the community.**

**Community Response to CCVs**
Borning and Muller suggest that the voice of the community should be more explicit in scientific reporting of value-sensitive studies [6]. Therefore, beyond including participant quotes, we also received feedback from the broader Wikipedia community at our Village Pump posting [66]. Editors affirmed that our five CCVs are in alignment with their perception of the community. They also emphasized that Human Authority is absolutely critical, and they appreciate P1's intention to maintain power in the hands of editors. Finally, they asserted that Positive Engagement and Community Trust are important, but also highly sensitive areas for the community. Any proposed means of enacting these particular CCVs should take care to substantially involve the community and ensure that the means are likely to have the intended effect.

**DISCUSSION**

**The Scope of the "Convergent Community Values"**
How generalizable are the five CCVs presented in this work? From a theoretical perspective, we follow Borning & Muller to take a humble approach, and thus are cautious about making claims of generality [6]. However, Borning & Muller also suggest a path forward: building up *"collections of case studies, heuristics (and yes, even lists) that are particularly relevant for a broad range of cultures and contexts."*

As a first step in this direction, we have identified studies that examined the role of "moderation" in various contexts, including the Turkopticon community of Amazon Mechanical Turk workers [39] and an assortment of communities organized as Reddit subreddits [7, 42], or Twitch channels and Facebook groups [52]. The moderation task in these communities is directly analogous to the Wikipedia quality control task that ORES supports. While not framed explicitly in terms of community values, these studies also found strong commitments to Positive Engagement, Workflow Support, Effort Reduction, and especially Human Authority in their contexts. In particular, quality control is important, but so are its effects on community participants. Both further literature review and studies can help establish the range of the contexts and communities where our CCVs are relevant.

**Integrating Values into Algorithmic System Design**
Algorithmic systems are designed and deployed at multiple levels. At the most basic, algorithmic systems are driven by an algorithm. In the case of ORES and other machine learning systems, this algorithm is a model that is trained on examples of labeled data. However, the algorithm ultimately provides only *signals*—often predictions at some confidence level. A User Interface (UI) is required to connect these algorithmic signals to a user's experience, and the design of this UI can affect how people make sense of these signals. At the highest level, there is a work process (i.e., "decision system" [47]) that has been designed to serve some common community need. Throughout results, we have included bolded recommendations for ways that each CCV can be operationalized; here, we posit that these recommendations should be integrated at all three of the *algorithm*, *UI*, and *work process* levels.

We will use Human Authority and Effort Reduction to provide concrete examples of how CCVs can guide system design in the vandalism fighting context. Protecting Wikipedia from vandalism involves ORES models, different UIs, and a set of work processes. Specifically, ORES provides a signal that is used to highlight "likely damaging edits". Various UIs use ORES predictions, e.g., focusing editors on discovering and removing vandalism as quickly as possible. Finally, a set of work processes have evolved around patrolling new changes as they are saved, with the help of these UIs.

*Example: Integration at the Level of the Algorithmic Model*
Applying the concept that human judgement should be the final arbiter on individual instances, one way to incorporate Human Authority at the algorithm/model level is to grant humans the choice to select specific model instances. We can design ways to allow tool developers and community members to search through system metrics to choose models that match their own operational concerns and needs. For instance, we can develop a syntax for requesting an optimization of certain system criteria from ORES—e.g. *minimum false-negative* can probably lead to a useful model for an automated counter-vandalism bot; *minimum false-positive* provides a useful model for semi-automated edit review; and *minimum false-negative-for-newcomers & overall-error < 0.1* yields a model that prioritizes newcomer protection in quality control.

*Example: Integration at the Level of the UI*
One challenge in applying the Human Authority CCV is that the humans who need to make final judgments (e.g., human patrollers in the context of counter-vandalism) might lack the necessary knowledge to understand and directly interact with algorithmic outputs. For example, ORES makes prediction errors, and understanding those errors is critical for human patrollers to appropriately respond to ORES' outputs. Designing an interactive visualization interface can be an effective approach to bridge literacy gaps by helping human patrollers to understand how the inputs of the algorithmic system relate to its outputs, explore algorithmic predictions and errors, and better work with algorithmic tools.

*Example: Integration at the Level of the Work Process*
Furthermore, various UIs and work processes should be designed to work together to achieve the best outcome, e.g., *both* Effort Reduction and Human Authority. For example, in counter-vandalism, an automated bot can serve as the first line of defense and automatically revert edits that are highly likely to be damaging; human patrollers can use the semi-automated tool to review and tag edits; and a socialization tool can review edits tagged by human patrollers to identify those that are not "damaging" based on a model that prioritizes newcomer protection, and then act accordingly (e.g. providing guidance to newcomers on how to improve their edits).

**Future Work on Managing Trade-offs**
*Develop Tools to Explain Value Trade-offs*
Our findings suggest a few sets of trade-offs in enacting different CCVs. For example, if an edit is likely good faith but very damaging, what is the cost of leaving this damaging edit live for a given amount of time versus the benefit of not discouraging a good faith editor? When Effort Reduction conflicts with Positive Engagement, what is the cost of requiring extra review work for moderators versus the benefit of giving content editors some additional "benefit of the doubt"?

One future direction is to develop methods and tools for community stakeholders to interrogate and navigate trade-offs. For example, we can develop methods to communicate machine learning models' value trade-offs to stakeholders. One possible way is to generate a value report for each model, with statistics of the system criteria corresponding to different community values and explanations of how different values are honored in this model. We can further generate visualizations to illustrate the relationship between different values.

*Design Solutions to Negotiate and Balance Value Trade-offs*
ORES is not a standalone machine learning model, but is embedded in Wikipedia—a rich and complicated socio-technical system. Therefore, we cannot address value trade-offs in ORES by considering only the machine learning model. Instead, we must expand the design space to include social structures along with the technology and algorithms. For example, the Wikipedia community has already developed numerous formal and informal mechanisms for resolving conflicts in content-related disputes and user conduct disputes, such as Third Opinion, Dispute Resolution Noticeboard, Mediation Committee, Request for Mediation, and Request for Comments. We can leverage, customize, and improve these existing consensus-building mechanisms to facilitate stakeholders to discuss and negotiate value trade-offs in ORES.

**ACKNOWLEDGMENTS**
We thank our anonymous reviewers, Zhiwei Steven Wu, colleagues at GroupLens Research at the University of Minnesota, and the HCI Institute at Carnegie Mellon University for their feedback. This work was supported by the National Science Foundation (NSF) under Award No. IIS-2001851 and IIS-2000782, and the NSF Program on Fairness in AI in collaboration with Amazon under Award No. IIS-1939606.